\documentclass[twocolumn,         
               showpacs,longbibliography,            
               showkeys,preprintnumbers,     
               aps,                 
               prd,          	    
               letterpaper,             
               nofootinbib,         
               tightenlines,        
               floats,floatfix      
               ]{revtex4-1}
\usepackage{physics}
\usepackage{epsf}
\usepackage{graphicx,color}
\usepackage{subfigure}
\usepackage{latexsym}
\usepackage{amsmath,amssymb}        
\usepackage[colorlinks=true,linkcolor=blue,citecolor=blue]{hyperref}
\usepackage{mathrsfs}
\usepackage{comment}
\usepackage{soul}
\usepackage{cancel}
\definecolor{purple}{rgb}{0.58,0.0,0.83}
\definecolor{orange}{rgb}{1,0.5,0}
\DeclareSymbolFontAlphabet{\mathrsfs}{rsfs}
\DeclareMathAlphabet{\mathcal}{OMS}{cmsy}{m}{n}

\begin{document}


\title{Black Holes as Condensation Points of Fuzzy Dark Matter Cores}


\author{Curicaveri Palomares-Ch\'avez}
\email{curicaveri.palomares@umich.mx}
\affiliation{Instituto de F\'{\i}sica y Matem\'{a}ticas, Universidad
              Michoacana de San Nicol\'as de Hidalgo. Edificio C-3, Cd.
              Universitaria, 58040 Morelia, Michoac\'{a}n,
              M\'{e}xico.}           

\author{Iv\'an  \'Alvarez-Rios}
\email{ivan.alvarez@umich.mx}
\affiliation{Instituto de F\'{\i}sica y Matem\'{a}ticas, Universidad
              Michoacana de San Nicol\'as de Hidalgo. Edificio C-3, Cd.
              Universitaria, 58040 Morelia, Michoac\'{a}n,
              M\'{e}xico.}               

\author{Francisco S. Guzm\'an}
\email{francisco.s.guzman@umich.mx}
\affiliation{Instituto de F\'{\i}sica y Matem\'{a}ticas, Universidad
              Michoacana de San Nicol\'as de Hidalgo. Edificio C-3, Cd.
              Universitaria, 58040 Morelia, Michoac\'{a}n,
              M\'{e}xico.}  


\date{\today}


\begin{abstract}
We simulate the formation of Fuzzy Dark Matter (FDM) cores in the presence of a Black Hole (BH) to explore whether BHs can serve as seeds for FDM core condensation. Our analysis is based on the core-condensation via the kinetic relaxation process for random initial conditions of the FDM. In a generic scenario the BH merges with a pre-collapsed mini-cluster formed in a random location, once they share location the core-condensation starts withe the FDM density centered at the black hole that during the process acquires a profile consistent with that of the stationary solution of the FDM+BH eigenvalue problem. These results indicate that BHs can indeed act as focal points for FDM core condensation. 
Furthermore, we find that the central density of the resulting FDM core depends on the mass of the BH, which due to its permanent motion relative to the FDM core during the evolution, produces a smaller core density for bigger BH masses; in this way the BH mass is a parameter leading to a new diversity of central FDM core densities. As a collateral result, for our analysis we revised the construction of stationary solutions of FDM+BH and found a phenomenological formula for the FDM density that can be used to fit FDM cores around BHs.
\end{abstract}




\maketitle


\section{Introduction}

Fuzzy Dark Matter (FDM) assumes dark matter is an ultralight boson with mass of order $10^{-23}-10^{-21}$eV, whose behavior differs from that of cold dark matter at galactic scales, in particular it promotes core formation and shows wave-like phenomena that produce a particular distribution of matter at galactic core and halo scales 
\cite{Chavanis2015,Niemeyer_2020,Hui:2021tkt,ElisaFerreira}. The structures formed by this particle are cores surrounded by envelopes with profiles similar to those of the CDM, the core being an essential fingerprint of the model  determined from structure formation simulations (e.g. 
\cite{Schive:2014dra,Mocz:2017wlg,Veltmaat_2018,MoczPRL2019,May_2021,Gotinga2022}).

Core formation has been studied in more detail at small scale simulations in basically two settings, one is the multi-merger of cores, for example in \cite{Schive:2014hza,Mocz:2017wlg,Schwabe:2016,periodicas,corehaloSR} that leads to a simple construction of core-halo structures. Another approach uses the kinetic relaxation as explanation and simulation of core formation out of random initial conditions \cite{Rusos2018,Chen2021}. These two methods help studying the core condensation process of bosonic configurations, their mass-growth and the envelope profile of their halos using small domain simulations \cite{Eggemeier2019,Chen2023,Purohit_2023,Chen2024}. 

An essential ingredient introduced in FDM phenomenology is the presence of black holes and their behavior within FDM cores. For example, in \cite{Chavanis_2019,Chavanis_2020} the FDM core density profile properties are studied under various regimes of the boson gas and scenarios that include a black hole, in \cite{Wang2022} the interaction of FDM and the BH is studied, in particular the dynamical friction and the drag of the wake behind the hole while moving, in \cite{Boey2024} also the dynamical friction of FDM cores in BH is studied in merger scenarios, in \cite{KooFPP2024} also the binary black hole merger moving inside FDM cores is studied in the context of the final parsec problem of the merger.
Other studies reveal the impact that FDM cores have at boosting supermassive black hole growth \cite{SchiveBHs}. In gravitational wave related contexts, the binary black hole within FDM is studied to estimate the extraction of angular momentum due to dynamical effects of clumps of dark matter within the core \cite{Bromley2024}, and more recently the scattering of FDM by the binary is analyzed \cite{tomaselli2025scatteringwavedarkmatter}.  In \cite{ElZant2020} the authors study the ejection of SMBHs due to the superposition of modes and the accumulation of random walk effects within FDM halos. Also in \cite{Lancaster_2020} the motion of a massive point particle in a FDM environment that includes its granules is studied, while in \cite{Boey2024} the collision between an FDM core and a SMBH is analyzed. These studies consider the black hole to be Newtonian, and focus mainly in dynamic effects of moving black holes. In a relativistic context, the coexistence and phenomenology of black hole with ultralight bosonic dark matter has also been analyzed, for example in 
\cite{Cardoso2022,Traykova2023,Ravanal2023,Boudon2023,Traykova2023} various effects of black hole dynamics on the scalar field are studied, including dynamical friction, while in \cite{Bamber2023,Aurrekoetxea2024} the analysis centers on the potential detection of ultralight dark matter via effects on gravitational wave phenomenology, including the case with self-interaction \cite{Aurrekoetxea_2024_selfinteracting}. In cosmological scenarios, studies include for example the nucleation of scalar clouds \cite{Hertzberg2020}.

In this work we study the effects of a BH during the FDM core-condensation, for which we follow the kinetic relaxation method developed in \cite{Rusos2018,Chen2021}, that uses random initial conditions, so that the granularity of the FDM distribution develops in the presence of the BH since initial time. In our analysis we use the Newtonian version of a spherically symmetric BH and add its gravitational effects to the Scr\"odinger-Poisson system of equations that rules the dynamics of the FDM+BH system, while we ignore the partial accretion of wave dark matter, which depends on the wave-length and thickness of wave-packets approaching the black hole, as demonstrated with full non-linear numerical relativity in \cite{GuzmanLora2013,BlinkingBlackHole}, as well as other non-linear accretion effects in non-symmetric scenarios \cite{Cardoso1}.

We find that the FDM distributes around the BH, where it starts the process of condensation and that its density approaches that of stationary solutions of the FDM+BH eigenvalue problem. In order to show this, we revise the construction of stationary solutions of this system following the ideas in \cite{Moczfdmbh} and build a phenomenological formula usable to fit FDM distributions during the evolution of the system. We also use these solutions to test our evolution code. 

During the condensation process we notice that the interaction of the FDM structure with the BH, that back reacts to the dynamics of the FDM, produces an oscillatory motion of the BH as described in \cite{Boey2024}, in this case in a granular non-smooth FDM distribution. We also find that the relative motion of the BH with respect to the core influences the final FDM distribution and flattens the central core density. We consider this effect is due to the BH scattering of FDM while moving within the core; in order to support this idea, we simulated the case in which the BH does not back-react to the gravity sourced by the FDM and observe that in such case there is no central density decrease.

The paper is organized as follows. In Section \ref{sec:eqs} we describe the set of equations used to simulate the system, along with a description and tests of our code. In Section \ref{sec:simulations} we describe the simulations carried out and their diagnostics. Finally, in Section \ref{sec:conclusions} we draw some conclusions of our analysis.
We include three appendices, one in which we revise the stationary solution of the FDM+BH eigenvalue problem and construct a phenomenological formula that mimics the FDM distribution; a second one in which we evolve the system without BH, that serves as a test of our code and for comparison with our findings; a third one in which we study the test-field regime that serves to understand better the role played by the BH motion.

\section{Equations and numerical methods}
\label{sec:eqs}

\subsection{Evolution Equations}

The system of equations that rule the dynamics of the FDM+BH system is the following Schr\"odinger-Poisson (SP) system of equations:

\begin{eqnarray}
    i\hbar\partial_t \Psi &=& -\dfrac{\hbar^2}{2m}\nabla^2\Psi + m V \Psi,\label{eq:schro}\\
    \nabla^2 V &=& 4\pi G (\rho_T - \Bar{\rho}_T)\label{eq:poissonV},\\
    \ddot{\Vec{x}}_{BH} &=& - \nabla V_{FDM},\label{eq:2aleyBH}\\
    \nabla^2 V_{FDM} &=& 4\pi G (\rho -\Bar{\rho}),\label{eq:poissonVFDM}
\end{eqnarray}

\noindent  where $m$ is the boson mass,  $\rho = m |\Psi|^2$ is the bosonic gas density, $\Bar{\rho}$ its mean density, $V$ is the gravitational potential due to the FDM and the BH, while $V_{FDM}$ is the potential only due to the FDM. We approximate the BH with a Gaussian density distribution:

 \begin{equation}
 \rho_{BH}=M_{MH}\delta(\Vec{x}-\Vec{x}_{BH})\simeq CM_{MH}e^{-|\Vec{x}-\Vec{x}_{BH}|^2/2\epsilon^2}
 \label{eq:bhmodel}
 \end{equation}
 
 \noindent  where $C$ is such that the integral of the distribution is the black hole mass $M_{BH}$. The total density of the system is $\rho_T=\rho+\rho_{BH}$ so that the BH contributes to the total gravitational potential of the system, and finally,  $\Vec{x}_{BH}$ is the position of the black hole.

In order to solve the SP system above, we use the  transformations $t = t_0 \Tilde{t},\Vec{x}=x_0\Tilde{\Vec{x}},V=V_0 \Tilde{V},\Psi= \Psi_0 \Tilde{\Psi}$ and $\rho=\rho_0\Tilde{\rho}$, where
$ t_0 = \frac{x_0^2 m }{\hbar}$,
$V_0 = \left(\frac{\hbar}{m x_0}\right)$, 
$\Psi_0 = \frac{\hbar}{\sqrt{4\pi G m^3}x_0^2}$,
$\rho_0 =\frac{\hbar^2}{4\pi G m^2x_0^4}$,
 that leave the SP system in dimensionless code units that only depend on the length scale parameter $x_0$. We solve the dimensionless problem using the code CAFE-FDM \cite{Alvarez_Rios_2022}, that implements a pseudo-spectral method to discretize spatial derivatives, an RK4 scheme for the evolution of the wave function, and the two Poisson equations are solved using the FFT method.  All simulations are carried out in a periodic cubic domain of side $L$ in code units, with resolution $\Delta x$.  For the BH distribution we use $\epsilon=0.1\Delta x$ because it gives the same results as in \cite{Boey2024}, where a BH is represented thorough its gravitational potential $V_\bullet = -\frac{G M_{BH}}{\sqrt{max(r^2,\epsilon^2)}}$ and used within Schr\"odinger equation to simulate scenarios with FDM and BH in isolated domains; in fact we define the BH in terms of the distribution $\rho_{BH}$ because it works in periodic domains as well.

\subsection{Initial conditions} 

The kinetic relaxation leading to core condensation uses random initial conditions for the bosonic gas, that eventually lead to the collapse of overdensities that in turn promote the bosonic gas condensation. The systematic study of this collapse process is understood from local simulations defined in \cite{Rusos2018,Chen2021}, where various distributions in the momentum domain are proposed. In our analysis we use a Gaussian distribution $\Psi(\Vec{p}) = Ae^{-p^2/2}e^{iS}$ in the momentum space, with $S$ a random phase in the range $[0,2\pi]$ at each point of the momentum space, where $A$ is a normalization factor. In order to validate our methods and code, we run one of the simulation in \cite{Chen2021}, specifically on a box of side $L = 18$, with resolution $\Delta x = L/128$ and total FDM mass  in the numerical domain $M_{FDM}=M = 1005.3$ in code units as in \cite{Chen2021}. As an example in physical units, we use the scenario where cores of FDM are shown to boost the growth of supermassive black holes \cite{SchiveBHs}, there the boson mass is $m=10^{-22}$eV, and the average FDM density is $\bar{\rho}\sim 1.477\times 10^8$M${}_{\odot}/$kpc$^3$. The integral in our domain gives $M=2.29\times 10^{10}$M${}_{\odot}$, whereas the numerical resolution and domain size for such boson mass are $\Delta x=42$pc and $L = 5.3749$kpc. Under these conditions, the smallest black hole mass in our analysis below will be $M_{BH}=M/256=8.96\times 10^7$M${}_{\odot}$ and the most massive one $M_{BH}=M/32=7.168\times 10^8$M${}_{\odot}$. Another important part of information is the momentum space distribution, notice that the total mass of FDM in the domain defines the value of the normalization constant $A$ of the momentum space distribution. Moreover, we use a distribution of width $\sigma=1$, at one sigma radius $p=1$, the associated velocity is $\sim 65$km/s, which is cold. 

In what follows we use code units and use the numerical setting above as default in our simulations involving BHs. A final specification of initial conditions is that for the BH the initial condition consists in setting its location at the center of the domain.

\subsection{Tests.}

As a first test, we reproduce the core condensation of FDM with kinetic relaxation using the initial conditions described above, that leads to results consistent with those in \cite{Chen2021}. The test in contained in Appendix \ref{app:pureFDM} and also serves to compare the condensation process with the cases with a Black Hole.

As a second test of our implementation, we evolve models of the stationary solutions of the FDM+BH eigenvalue problem developed in \cite{Moczfdmbh} and revised in Appendix \ref{app:stationary}. We show that the solutions are stable and that our phenomenological formula is usable.

\section{Simulations and analysis}
\label{sec:simulations}

\subsection{Simulations}

\begin{figure}
    \centering
     \includegraphics[width=8.5cm]{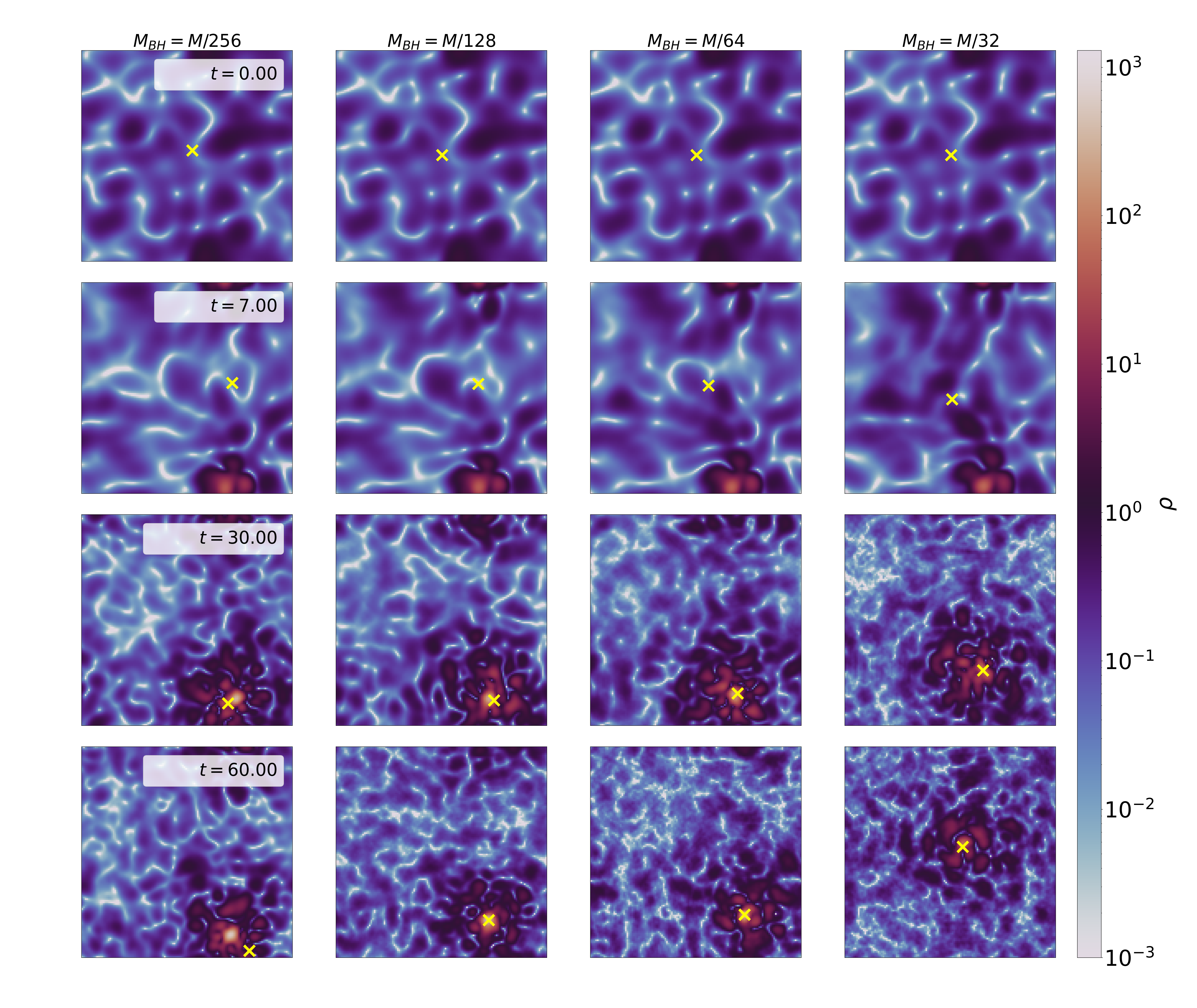}
    \caption{Snapshots of the density projected on a plane that contains the position of the black hole and  is parallel to the $xy-$plane of the numerical domain. The black hole position is represented with a yellow cross at different times for simulations with $M_{BH} = M/256, M/128, M/64$ and $M/32$. The evolution illustrates the formation of a minicluster near the bottom right corner. Later on the minicluster and the BH merge and start traveling together while the BH position oscillates with respect to the core, which can be seen in the movies in the supplementary material. By $t\sim 60$ the FDM approaches condensation as seen in the analysis below.}
    \label{fig:evolutionBR}
\end{figure}

We performed 32 simulations with the same type but different seed to generate random initial conditions for the FDM and various black hole masses $M_{BH}=M/256,M/128,M/64,M/32$, which lie within the range of masses used in studies of dynamical friction \cite{Wang2022} and those of PBHs used for nucleation of axion stars in \cite{Hertzberg2020}. 
Different initial seeds for initial conditions lead to different evolutions of the FDM in the presence of a BH, however the results of the simulations can be classified into two sets: a) a minicluster of FDM forms near the BH and its further collapse, relaxation and core condensation happen around the BH all the way, b) a minicluster of FDM forms  beyond the tidal radius of the BH and its further evolution includes the merger with the BH where the relaxation and condensation into a core happens. Among these two scenarios the second one is the non-trivial case that better shows how the BH acts as a condensation point for FDM density. In order to illustrate the evolution of this generic case, we use a particular seed to generate the random phase of FDM in Fourier space that in turn generates the initial conditions of the FDM, and study its evolution using the four black hole masses. The evolution of these four simulations is presented in Figure \ref{fig:evolutionBR}, that contains snapshots of the FDM density and the Black Hole position. At early times by $t\sim 7$ a minicluster forms near the bottom-right corner, later on this cluster merges with the black hole, following different trajectories for each black hole mass, and finally the condensation takes place at the black hole location. Animations of these simulations appear as Supplementary Material \cite{example2024}.

\subsection{Condensation.} 

\begin{figure}[h]
    \centering
    \includegraphics[width=8.5cm]{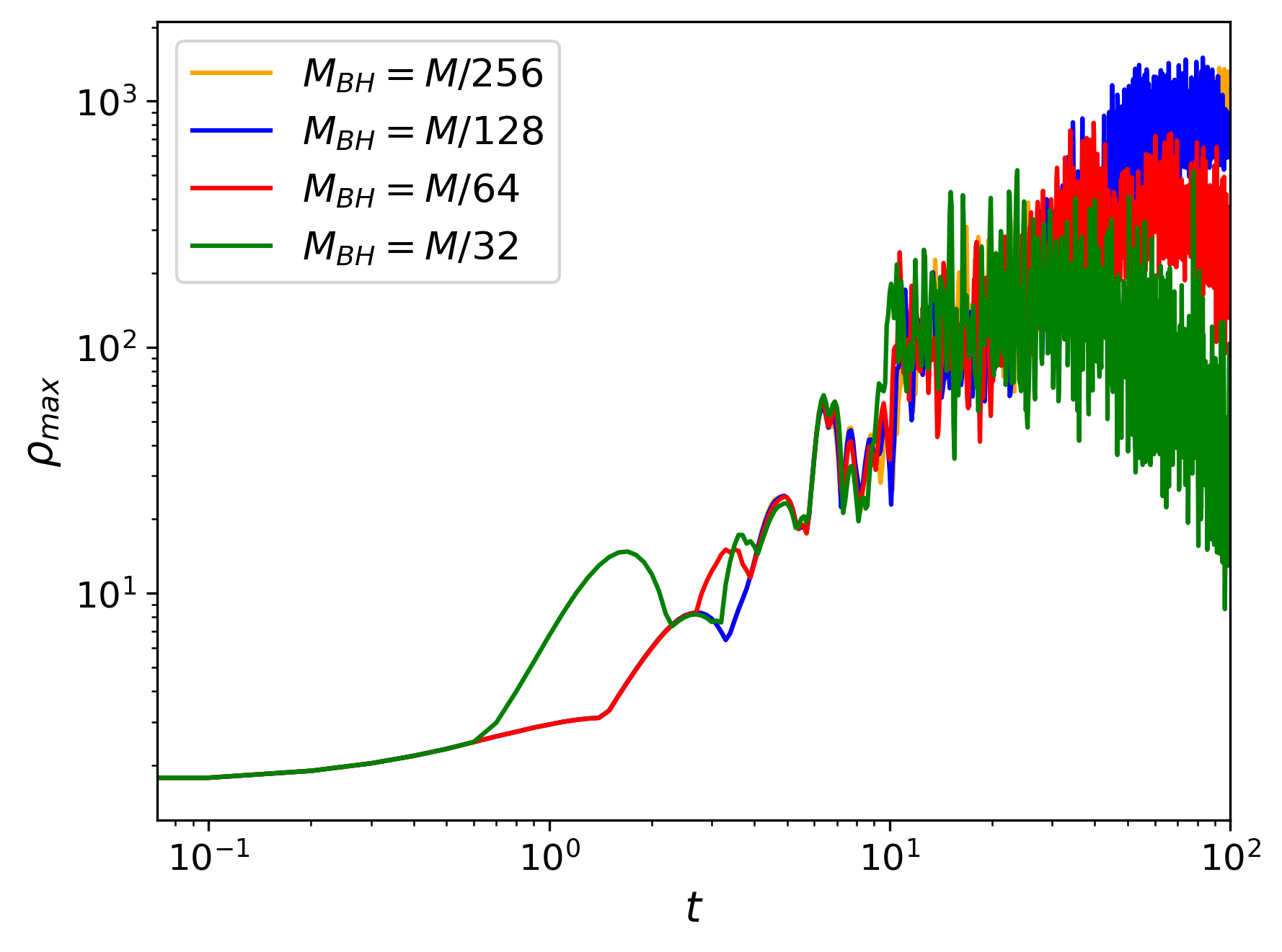}
    \caption{Evolution of $\rho_{max}$ for the various black hole masses $M_{BH}=M/256,M/128,M/64,M/32$. Notice that for the less massive BHs the density continues growing until $t\sim 60$ where it stabilizes.  A difference between small BH masses and the case $M_{BH}=0$ shown in Fig \ref{fig:rhomaxCondensation}, is that $\rho_{max}$ stabilizes with a BH, whereas it keeps growing when there is not BH in this time window, although as we will see below, the central density for the cases $M_{BH}=0,M/256$ and $M/128$ are very similar. For the most massive BHs the maximum density decreases, which is an indication of the effects of the presence of the BH. For comparison of the evolution without BH 
    see the simulations in Appendix \ref{app:pureFDM}.}
    \label{fig:rhomax_moving}
\end{figure}

In order to see if condensation happens, we monitor the evolution of $\rho_{max}$ and the results are shown in Figure \ref{fig:rhomax_moving}, where we notice that the condensation takes place, the density grows with a power law of $t$ starting after $t\sim 30$, which indicates the beginning of condensation, however later on the maximum density decreases or stabilizes instead of growing up as in the pure FDM scenario in \cite{Rusos2018,Chen2021}.
The decrease of $\rho_{max}$ is smaller for smaller black hole masses and can be understood in terms of the relative motion between the BH and the core. That is, the black hole, by being in motion, drags the FDM around it and prevents the density from increasing, with the effect being stronger for larger black hole masses. 
Thus, the presence of the black hole has the effect of flattening the FDM core density in the long term.

We contrast this density behavior with that it the test field case developed in Appendix \ref{app:testfield}, in which the Black Hole does not back-react to the dynamics of the FDM and remains at a fixed position. In this scenario the FDM is attracted toward the hole and condensates around it, the black hole does not move, there is no re-heating due to the FDM dynamics and the BH does not scatter the FDM around. The comparison between Figures \ref{fig:rhomax_moving} and \ref{fig:rhomaxMBHtestfield} explains this effect.

\subsection{Relative motion.} 
\begin{figure}[h]
    \centering
    \includegraphics[width=7.5cm]{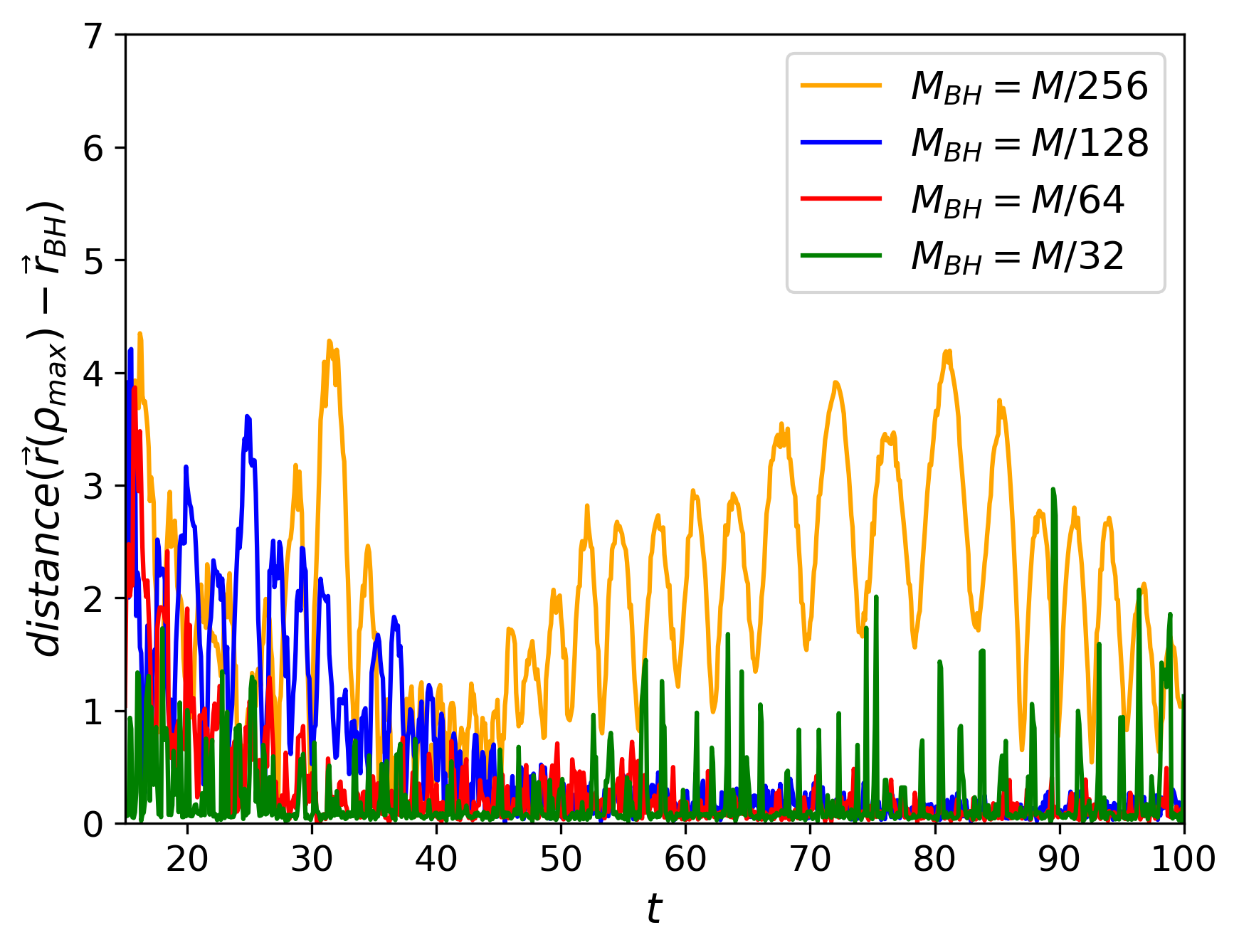}
        \caption{Distance from the location of maximum FDM density to the black hole for $M_{BH}=M/256,M/128,M/64,M/32$ as function of time. There is no trend in the relative distance as function of $M_{BH}$, because the BH in each simulation catches up with the minicluster at different times, with different velocities, at different points within the FDM granularity, so that heating and friction act differently in the different cases.} 
    \label{fig:r_t_moving}
\end{figure}

In Figure \ref{fig:r_t_moving} we show the distance from the point of maximum density to the Black Hole, for the four values of $M_{BH}$. The non-linear behavior and the dynamics of the granularity lead to a behavior that does not have a trend, since the motion of the black hole can be reheated differently since the different four mergers occur under different conditions, at different places and with different merging velocities, as can be seen in the animations and in Figure \ref{fig:evolutionBR}. 

\begin{figure}
    \centering
    \includegraphics[width=9cm]{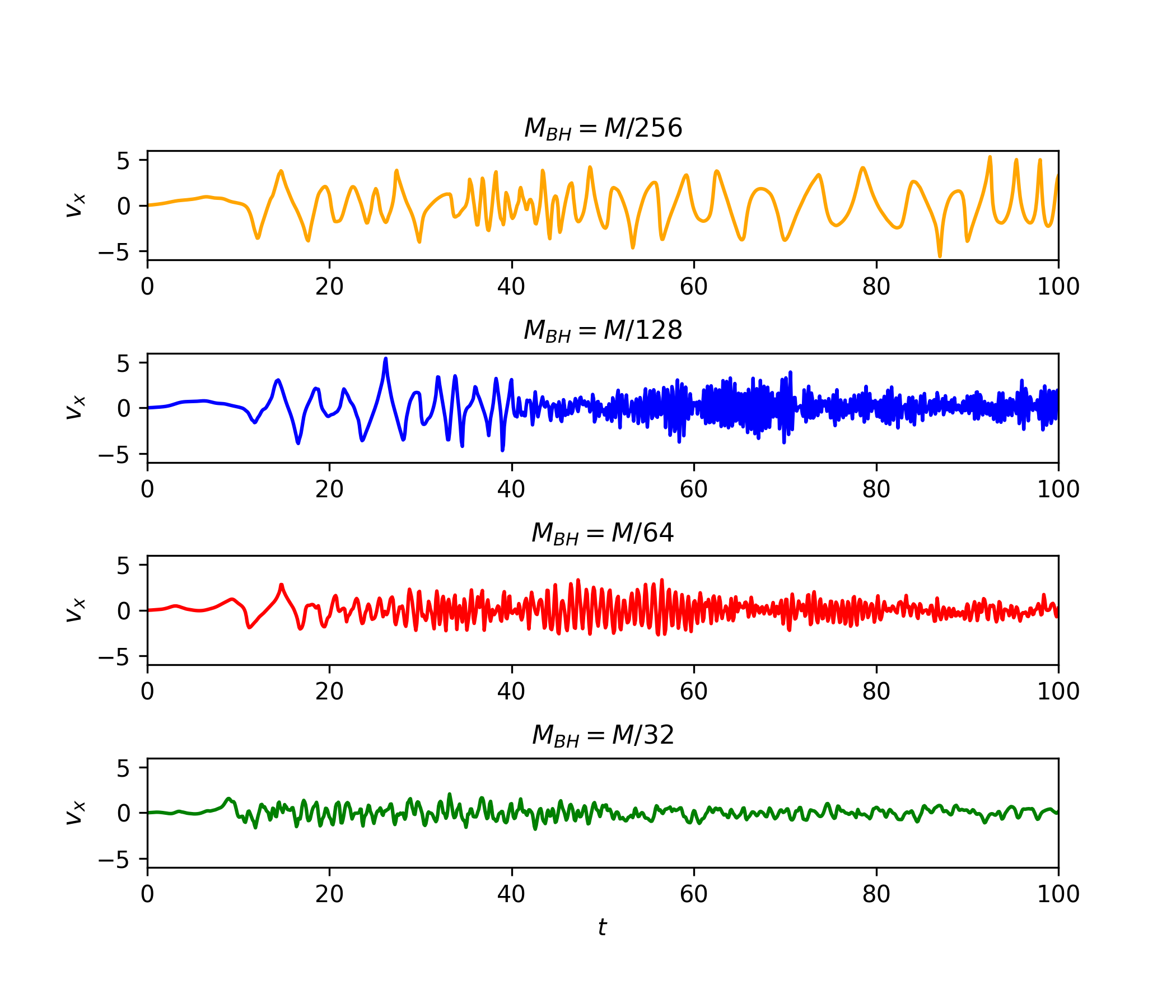}
        \caption{The $x-$component of the BH velocity for $M_{BH}=M/256/,M/128,M/64,M/32$. The oscillations are consistent with those proposed in \cite{Boey2024} for mergers of smooth cores and BHs. In our case the core has all its granules and time dependency since the beginning. Notice that for $M_{BH}=M/128,M/64$ the modulation of the velocity amplitude is consistent with the reheating of the BH motion. The behavior of the $y$ and $z-$components of the velocity is similar. The animations in the Supplementary Material illustrate the whole picture of the process.} 
    \label{fig:movingvxvyvz}
\end{figure}

Another part of the analysis is the velocity of the BH. In Figure \ref{fig:movingvxvyvz} we show the $x-$component of the black hole's velocity, which shows the oscillatory behavior studied under smoother FDM distributions in \cite{Boey2024}. Moreover,  for the cases $M_{BH}=M/128$ and $M/64$ the reheating pulses can be observed in the velocity as also described in \cite{Boey2024}.
Unlike in \cite{Boey2024} the oscillatiory motion is less regular because we do not assume an initially smooth spherically symmetric core, but deal with the core resulting from the collapse of FDM, with all its granular multipolar components and time dependence, since the formation time.

\subsection{Core fitting.} 

\begin{figure}[h]
    \centering
    \includegraphics[width=8cm]{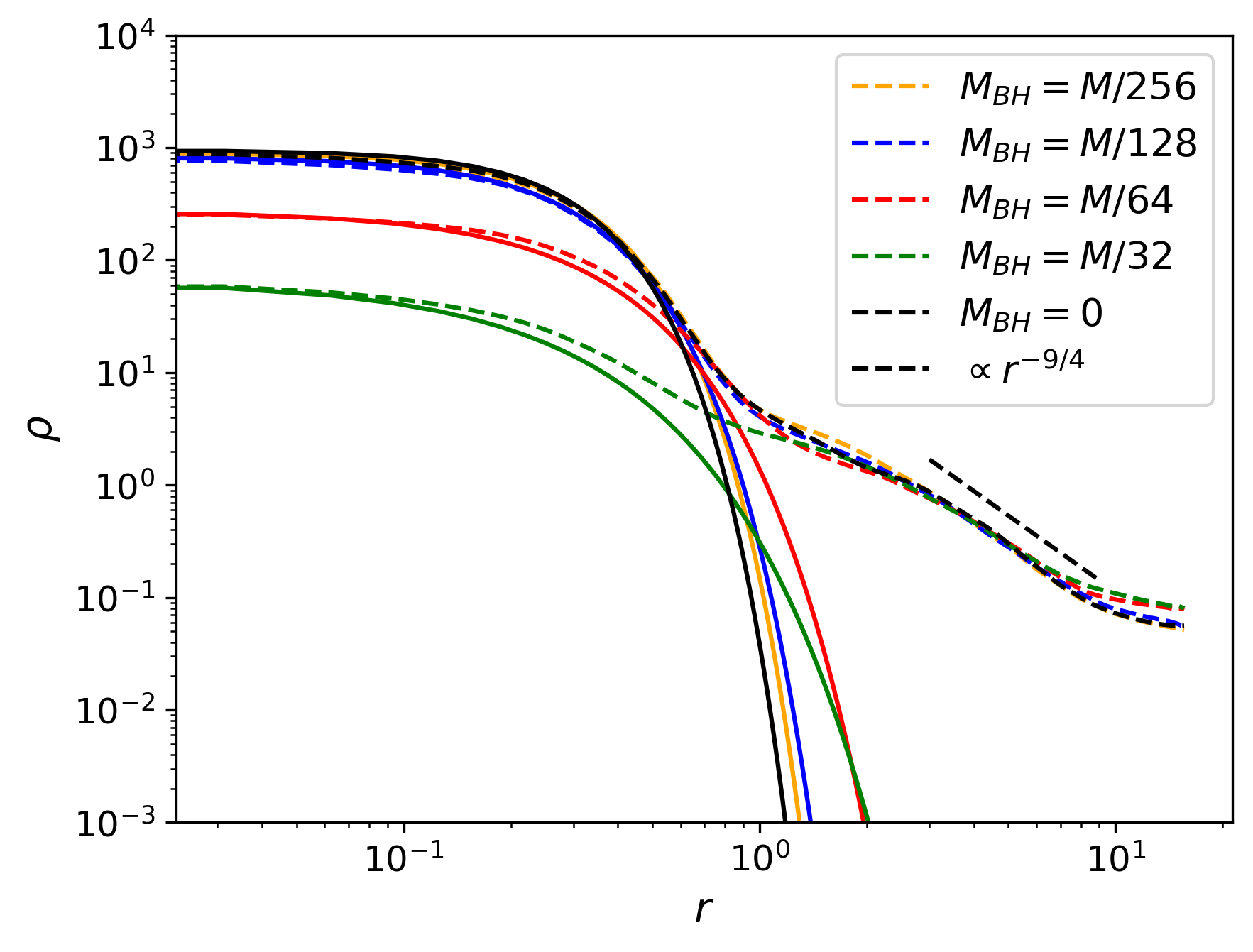}
    \caption{Density fitting of the angularly and time averaged density of FDM around the BH, using the phenomenological formula (\ref{rho_formula}), for the simulations with $M_{BH} = M/256, M/128, M/64, M/32$. The time interval  used to average density profiles is $t \in [70,100]$. Notice that the more massive the black holes, the smaller the central core density, which is consistent with the notion of a Black Hole moving within the core while dragging FDM around and preventing the density from growing, which is consistent with the results in Figure \ref{fig:rhomax_moving}. We enclose the case for $M_{BH}=0$ to show that when the mass of the BH approaches zero, the core approaches the case without BH.}
    \label{fig:time_average_30_80}
\end{figure}

Even though there is dynamics, we fit the resulting density profile of the FDM, not only at a fixed time but averaged over a time window using a phenomenological model for the density, that mimics the solution of the FDM-BH stationary eigen-problem in \cite{Moczfdmbh}, as described in Appendix \ref{app:stationary}; the FDM density fitting uses formulas (\ref{rho_formula})-(\ref{eq:betamodel}). The result shows the possibility to fit the configuration with an equilibrium profile of the FDM-BH eigen-problem. The fittings are presented in Figure  \ref{fig:time_average_30_80}, that we obtain from averaging the density profile in the time window $t \in [70,100]$, which is a lapse when, according to Fig. \ref{fig:rhomax_moving}, the maximum density profile ends growing during condensation for the smallest black holes. Notice that the central core density is smaller for bigger BH mass, whereas the central core density converges to the case without BH when $M_{BH}$ becomes smaller, in fact the cases $M_{BH}=M/256$ and $M/128$ have a profile similar already to that of the case $M_{BH}=0$, which is an interesting  consistency check of the simulations. This does not mean that small BHs play no role at all, notice that for all BH masses the minicluster truly moves and merges with the BH; this shows that even if small, BHs still play the role of condensation points.

\section{Conclusions} 
\label{sec:conclusions}

From our simulations we find that prior to the core condensation, miniclusters of FDM can form, that would collapse in themselves and condensate as shown in \cite{Rusos2018,Chen2021} and in Appendix \ref{app:pureFDM}. However the presence of the BH modifies such evolution and the mini-cluster merges with the black hole leading to the following  effects. First, a rather expected result between two attracting objects, the FDM gas density distributes around the black hole, despite the precollapsed minicluster is beyond the BH's tidal radius or not, which shows how the black hole acts as an attraction point even for the smallest black hole mass case used. 

Second, the motion of the BH with respect to the center of the core is oscillatory, a behavior found in \cite{Boey2024} while studying the dynamical friction of FDM on black holes using formed spherically symmetric cores; here we showed that the oscillatory motion happens even prior to core-condensation, within a non-spherically symmetric, time-dependent granular FDM distribution. The reheating of the BH motion due to the dynamics of the FDM can be observed as well.

Third, the analysis of $\rho_{max}$ as function of time reveals that the FDM cloud starts condensing once it is around the BH, however its growth is affected by the presence of the BH motion. The motion of the BH within the core disperses the central density away, reducing the averaged central density of the FDM core. We find smaller central density of FDM for a bigger BH mass, which can be understood in terms of the relative motion of the BH with respect to the core. This effect is contrasted with the test-field case, in which the BH does not back-react to the gravity of the FDM and does not scatter the FDM around.

Fourth, we have developed a spherically symmetric phenomenological formula for the FDM density that can be used to fit FDM cores around BHs. The accuracy of this model suggests that stationary solutions act like attractor solutions of FDM in the presence of a Black Hole.

In summary, our analysis shows that Black Holes act as condensation points where kinetic relaxation can take place. We consider that the influence of the black hole in the galaxy formation context can lead to consistency checks of the FDM model, particularly in scenarios of galactic cores with supermassive black holes at their centers. The effects described in this work can lead to new detectable predictions that in turn should challenge the viability of the FDM model.
Potential scenarios to be analyzed include the motion of the BH relative to the core, for example, the oscillations seen in Fig. 3 for the smallest black hole have a period of order 14Myr, that eventually could be correlated to the variability  near SMBHs, and its frequency should be correlated with the granularity of the FDM distribution for a statistically significative  number of initial conditions. In this same context heating and reheating when the BH mass is small, near the test particle regime, can lead to chaotic motion within FDM cores \cite{alvarezrios2025unveilingorbitalchaoswild} and in turn contribute with wandering BHs within cores. Another potential application is related to the evidence that the presence of the BH affects the averaged central FDM density as seen in Fig. 2, which allows the comparison with the systematic modeling of dark matter distribution at M87 \cite{2022ApJ...929...17D}, using also luminous matter models coupled to FDM \cite{alvarezrios2024fermionbosonstarsattractorsfuzzy}.


\section*{Acknowledgments}
We thank Pierre-Henri Chavanis for providing useful comments. Curicaveri Palomares and Iv\'an \'Alvarez receive support from the CONAHCyT graduate scholarship program. This research is supported by grants CIC-UMSNH-4.9, Laboratorio Nacional de C\'omputo de Alto Desempe\~no Grant Nos. 1-2024 and 5-2025, CONAHCyT Ciencia de Frontera 2019 Grant No. Sinergias/304001.


\bibliography{BECDM}

\appendix

\section{Pure FDM}
\label{app:pureFDM}

\begin{figure}
    \centering
    \includegraphics[width=8cm]{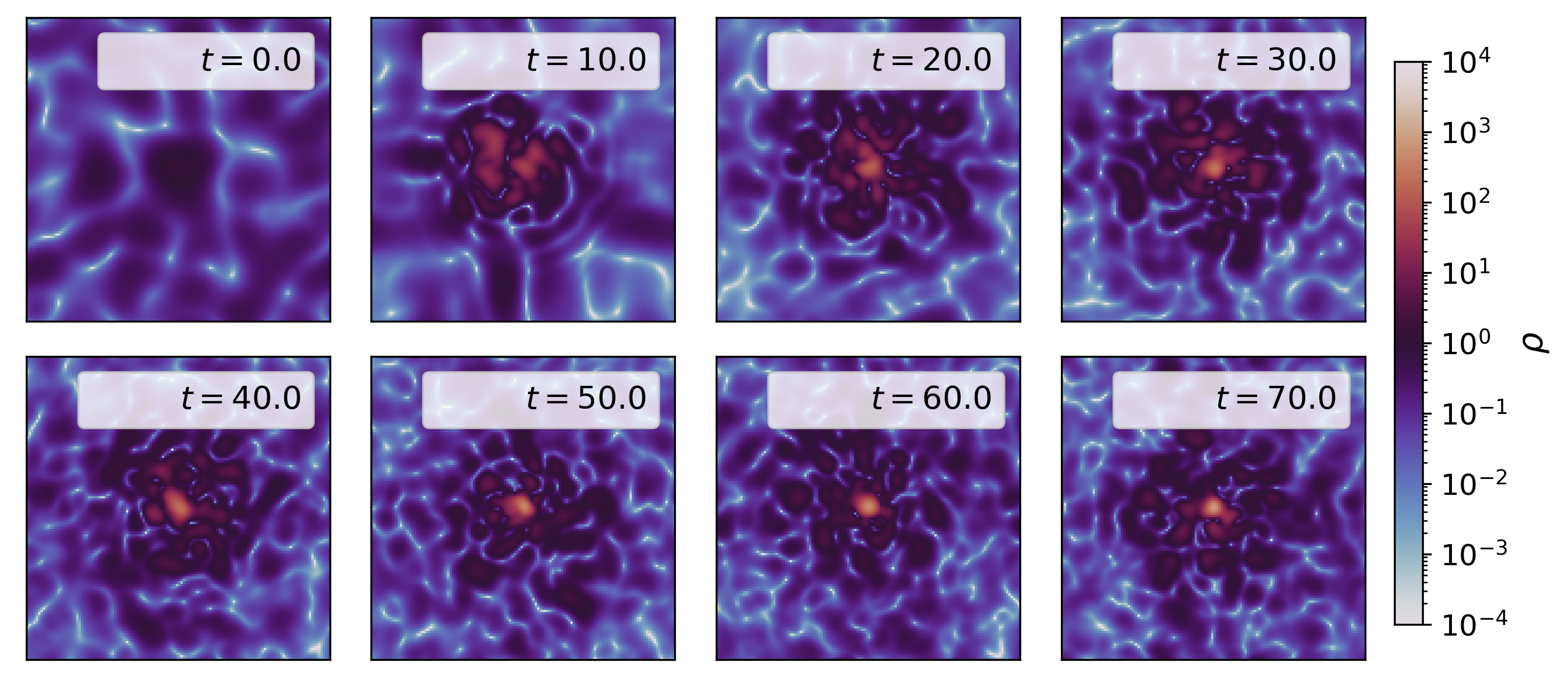}
    \caption{Snapshots of the density on a plane that passes through the maximum density at various times. This simulation illustrates the condensation of an FDM core, resembling the results in \cite{Chen2021}. }
    \label{fig:density_evolution}
\end{figure}

As a test we show the core condensation without a BH in Figure \ref{fig:density_evolution}, that contains snapshots of the density at different times on a plane that passes through  the point of maximum density, where we can notice that between $t \approx 10$ and $t \approx 30$ a minicluster is formed and in later times the solitonic core gets condensed. 

\begin{figure}[h]
    \centering
    \includegraphics[width=8cm]{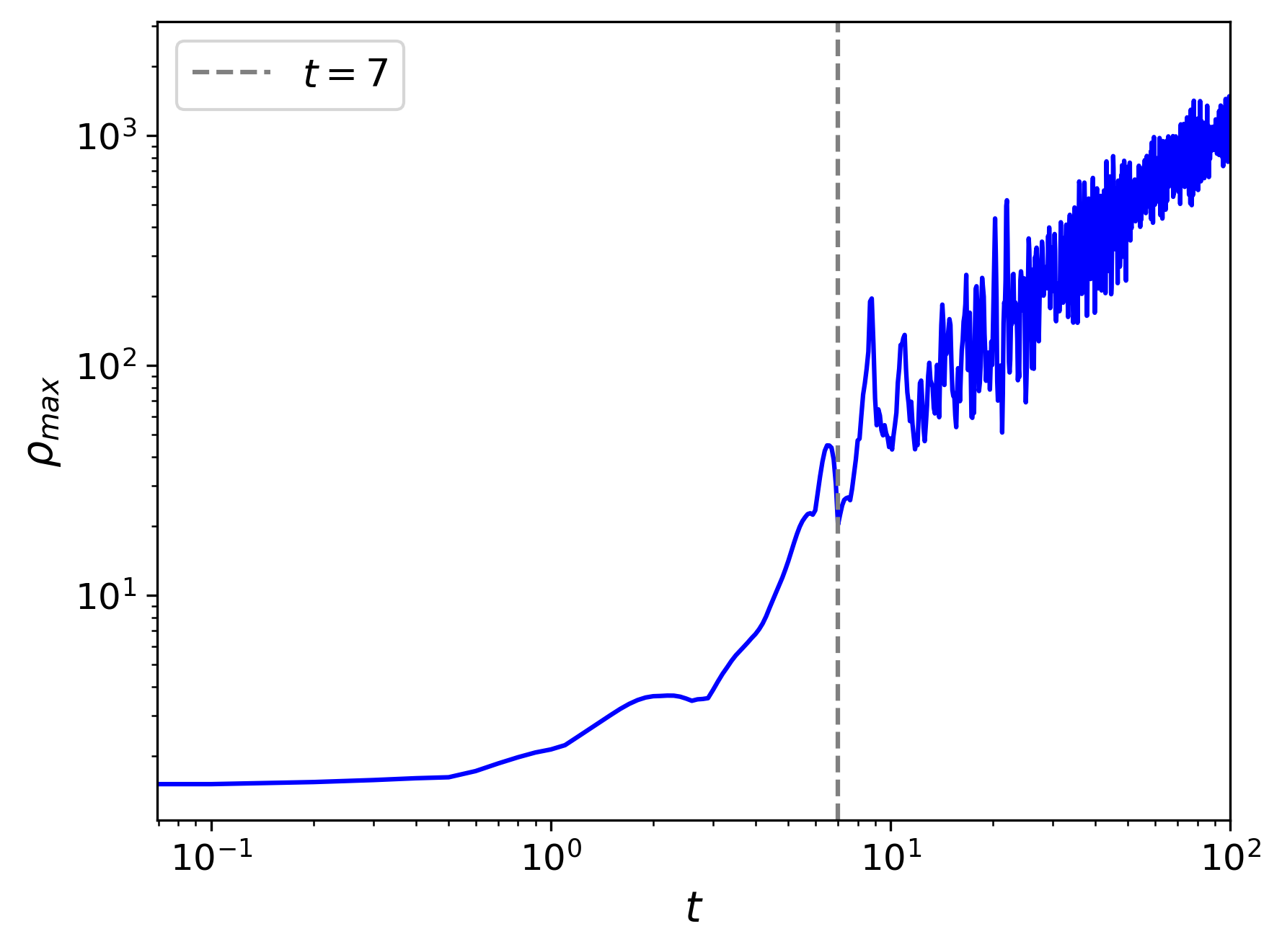}
    \caption{Maximum density $\rho_{max}$ in the numerical domain as function of time. This plot illustrates the condensation process as shown in \cite{Rusos2018}, which establishes the time scale $\tau_g$ at $t \sim 7$ for condensation, indicated with the vertical line.}
    \label{fig:rhomaxCondensation}
\end{figure}

The initial conditions in Fourier space and integrated mass, correspond to a standardized FDM set up in \cite{Chen2021}; we use this Appendix to verify that our code reproduces the results of such standard simulation. Due to the randomness of the initial conditions, the collapse of the minicluster s the result of a random fluctuation in a random location of the domain, where the configuration later collapses. The condensation process appears in Figure \ref{fig:rhomaxCondensation} where the maximum density is plotted as function of time in units of the condensation time  $\tau_g \propto \frac{v^6}{\Bar{\rho}^2 \log{(v L)} }$ as found in \cite{Chen2021}, which for our simulation corresponds to $\tau_g \approx 7$. The results obtained are in agreement with the findings in \cite{Chen2021} related to density growing with a power law of $t$.

What is shown in our analysis is that when a BH is introduced within the FDM initial conditions, the condensation process develops at the BH, even if the minicluster formed in a random location has to be first disrupted and the redistributed around the BH.

\section{Stationary Solutions of the FDM+BH eigenvalue problem}
\label{app:stationary}


We construct stationary solutions for the FDM+BH problem following \cite{Moczfdmbh}, whose equations are equivalent to the system (\ref{eq:schro})-(\ref{eq:poissonVFDM}) in the stationary regime 

\begin{eqnarray}
    i\hbar\partial_t \Psi &=& -\dfrac{\hbar^2}{2m}\nabla^2\Psi + m_B \left(V+V_\bullet\right) \Psi,
    \label{eq: GP}\\
    \nabla^2 V &=& 4\pi G (\rho - \Bar{\rho})\label{eq:poissonSP},
\end{eqnarray}

\noindent where $V_\bullet = -G M_{BH}/r$ is the gravitational potential due to a black hole of mass $M_{BH}$. The problem is solved in code units, spherical symmetry and harmonic time dependence of the parameter order  $\Psi = \psi(r)e^{-iwt}$ with $\psi$ a real function. These assumptions lead to the following eigenvalue problem for the eigenvalue $\omega$, provided isolation boundary conditions at infinity and regularity at the origin:

\begin{eqnarray}
    w \psi &=& \left[-\dfrac{1}{2r^2}\frac{d}{d r}\left(r^2\frac{d}{d r}\right) + \left(V+V_\bullet\right) \right]\psi,
    \label{eq: sphericalPsi}\\
 &&   \frac{1}{r^2} \frac{d}{d r}\left(r^2\frac{d}{d r}\right) V = \left(\rho - \Bar{\rho}\right).\label{eq:sphericalPoison}
\end{eqnarray}

\noindent We rewrite this problem with a set of two first order equations by defining the variables $\phi = r^2\psi'$ and $M = r^2 V'$, where $'$ denotes derivative along the radial direction. The eigenvalue problem as aa first order system reads:

\begin{eqnarray}
    \frac{d \psi}{d r} &=& 2r^2(-w+V+V_\bullet)\psi,\nonumber\\
    \frac{d M}{d r} &=& r^2(\rho - \Bar{\rho}),\nonumber\\
    \frac{d \psi}{d r}\psi &=& \frac{\phi}{r^2},\nonumber\\
    \frac{d V}{d r} &=& \frac{M}{r^2},\label{eq:eigenVP}
\end{eqnarray}

\noindent that we solve  using the shooting method  with the boundary conditions at the origin $\psi(0) = 0,M(0)=0,\psi(0) = \psi_0, V(0) =V_0$, and at an external boundary $\psi(r\rightarrow\infty) = 0,\phi(r\rightarrow\infty)=0$ and $V(r\rightarrow\infty)=0$, being $\psi_0$ a fixed central value; then the potential is rescaled to impose a monopolar boundary condition at the external boundary.

\begin{figure}
    \centering
    \includegraphics[width=8cm]{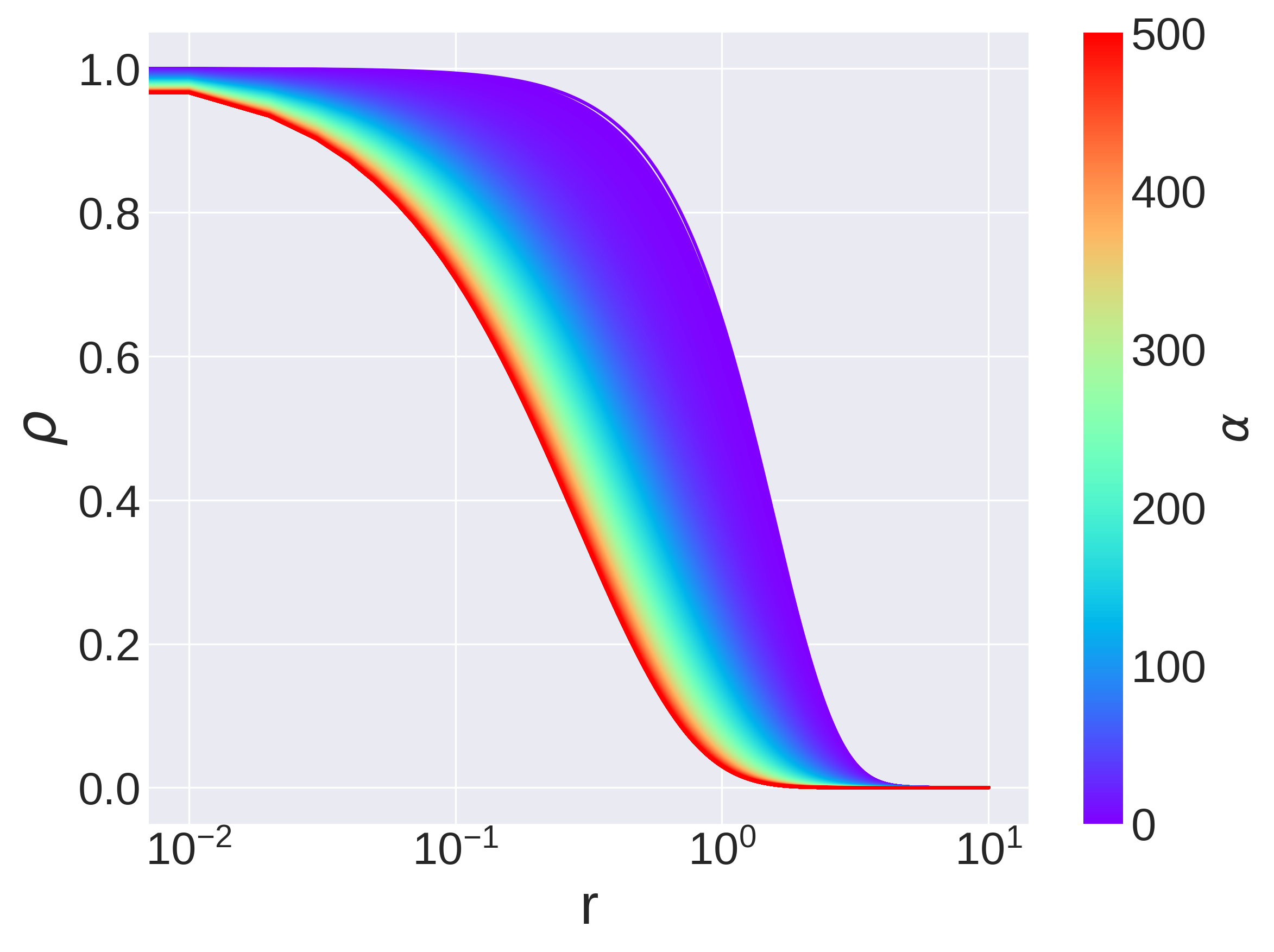}
    \caption{Solution densities obtained from the numerical solution of the eigen-value problem (\ref{eq:eigenVP}) for  \(\alpha\) in the range \([0, 500]\). The figure shows that the FDM density  becomes more compact as \(\alpha\) increases.}
    \label{fig:rho_solution}
\end{figure}

We note that this system exhibits invariance under the transformation
$\{t',\Psi',r',V',\rho',M_{BH}'\}  ~\rightarrow 
\{\lambda^{-2} t, \lambda^2 \Psi, \lambda^{-1}r, \lambda^2V, \lambda^{4}\rho,  \lambda M_{BH} \}$,where $\lambda$ is an arbitrary parameter. Now, for the construction of a phenomenological model of the FDM density we define the invariant $\alpha = M_{BH}^2/\psi_0$ under the $\lambda-$transformation, that allows one to parametrize the  family of solutions of the eigenvalue problem. In order to cover a wide parameter space, we solve the system  for values of $\alpha$ within the range $\alpha \in [0,500]$, and the results are shown in Figure \ref{fig:rho_solution}, where the $\alpha = 0$ is the ground state soliton solution for the SP system corresponding to the case without black hole.

We note that in the limit $V_{\bullet} \gg V$ or equivalently $\alpha \gg 1$, the system reduces to the case of the hydrogen atom  with  ground state solution $\psi(r) = 2 \left(\frac{1}{a_0}\right)^{3/2} e^{-r/a_0}$, being $a_0$ the Bohr radius. This exponential behavior can be seen in the numerical solution for high $\alpha$ in Figure \ref{fig:rho_solution}, thus we propose a formula that in the limit of high $\alpha$ approaches the exponential solution

\begin{equation}\label{rho_formula}
    \rho(r,\alpha) = \rho_c e^{-\ln2\left(\frac{r}{r_c}\right)^\beta},
\end{equation}

\noindent where $\rho_c$ is a central density, a core radius $r_c$ is defined as the radius where the density decreases to the half of its central value and $\beta$ is an $\alpha$, dependent function to be found.

\begin{figure}[h]
    \centering
    \includegraphics[width=4cm]{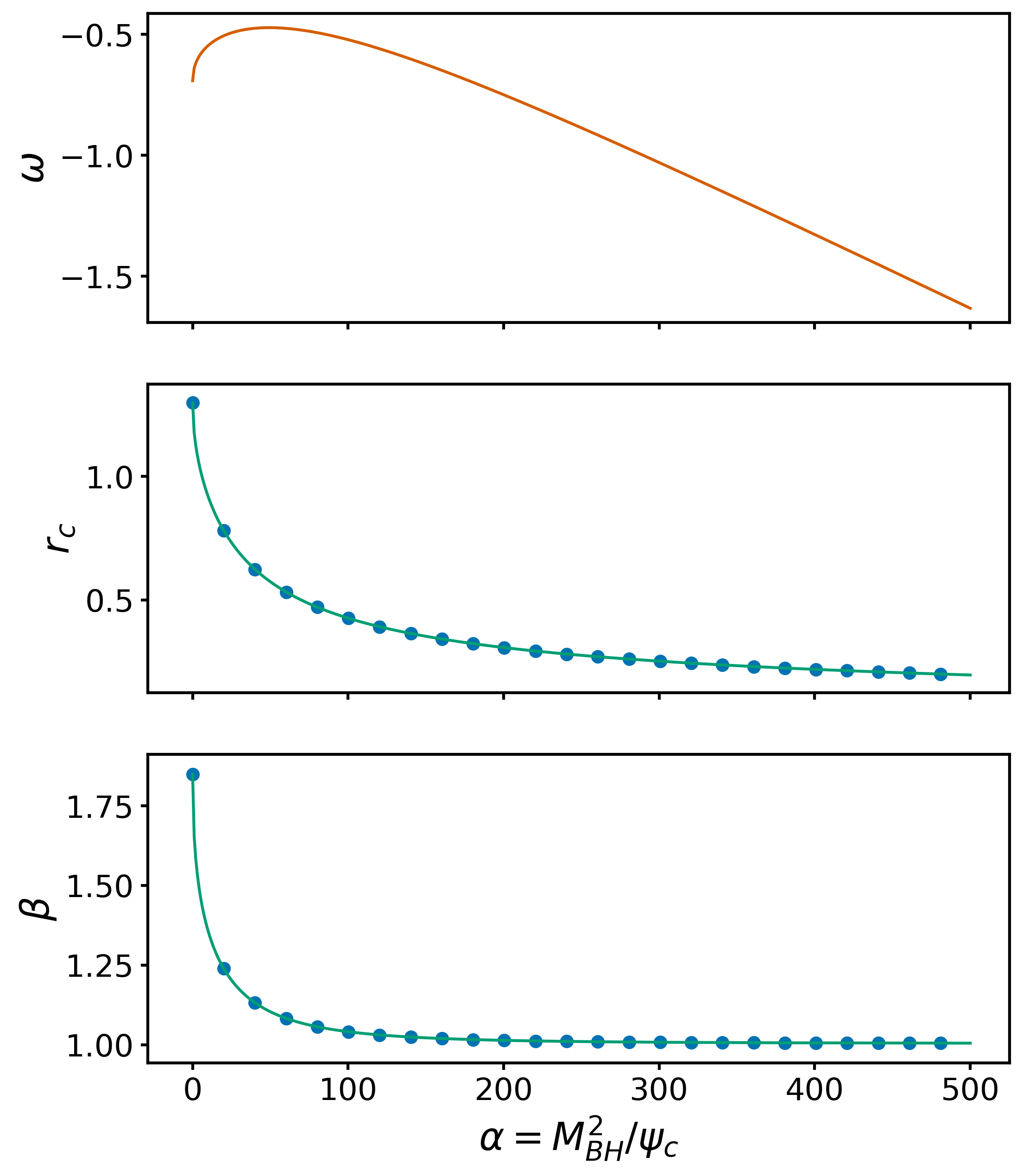}
    \includegraphics[width=4cm]{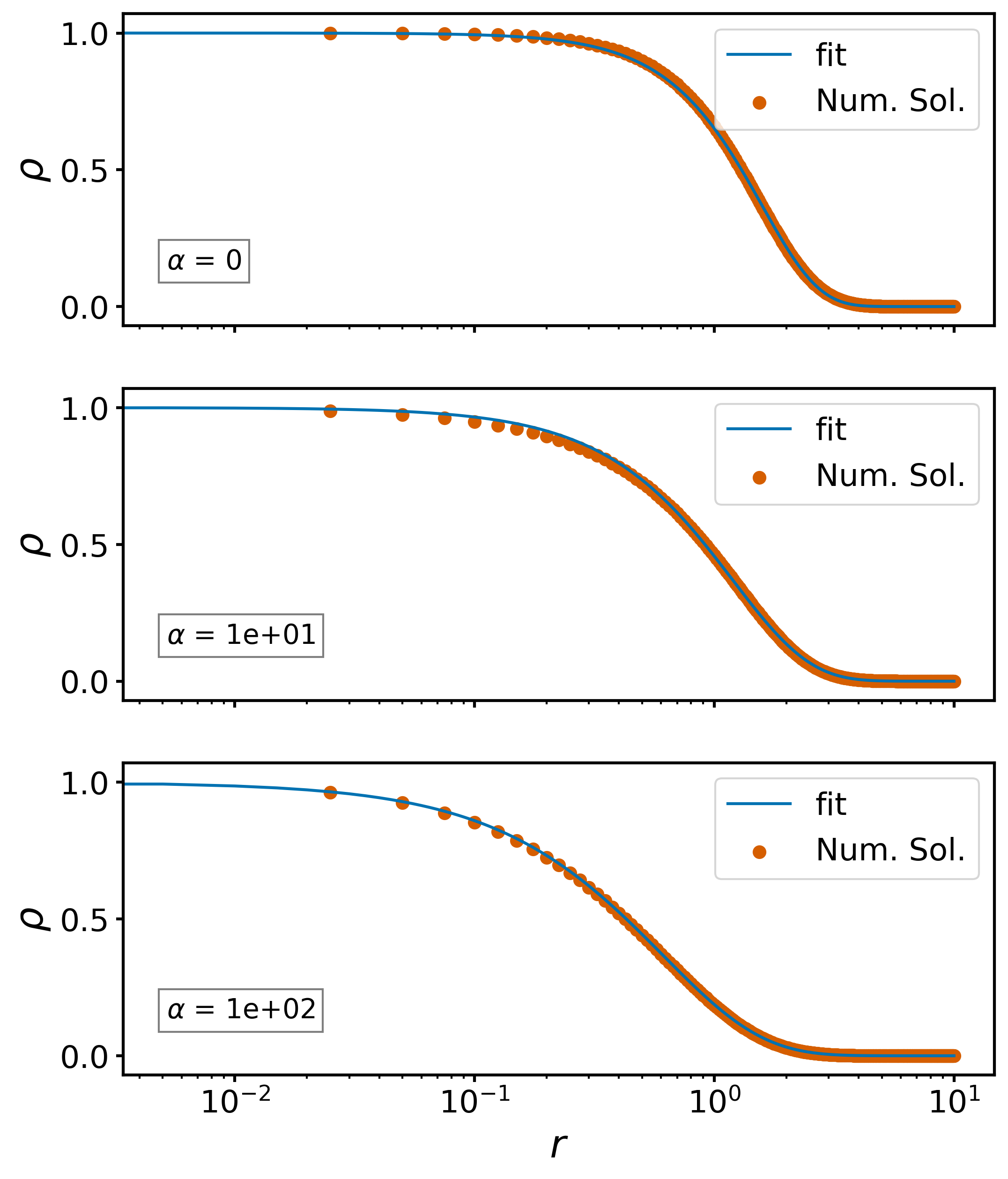}
    \caption{(Left) At the top we show the eigenvalue $\omega$ as function of $\alpha$. In the middle and bottom we show  $r_c$ and $\beta$ as functions of $\alpha$, where the dots correspond to genuine solutions of the eigenvalue problem and the continuous lines correspond to the values with our model formulas (\ref{eq:rcmodel}) and (\ref{eq:betamodel}). (Right) Numerical solution of the eigenvalue problem, together with the density resulting from formulas  (\ref{rho_formula}), (\ref{eq:rcmodel}) and (\ref{eq:betamodel}) that model the solutions. In this plot we use $\alpha = 0,10,100$, in order to show that the model works fine for configurations with $\alpha$ orders of magnitude different.}
    \label{fig:rc_beta_fit}
\end{figure}

In pure FDM scenarios, the ground state solution of the SP system (e.g. \cite{GuzmanUrena2004}) is difficult to use in phenomenological fitting of structures that are evolving within a simulation, then a practical formula por density was proposed to universally model FDM cores useful \cite{Schive:2014dra}. Here we do something analogous, since the solutions of the FDM+BH eigenvalue problem cannot be used to monitor the core formation around a black hole, thus we need a practical density profile that can be used during a simulation to fit the core density of the eigen-solution. To do so, 
we search for a function that can fit the core properties, for example we choose for the core radius $r_c$ the function

\begin{equation}
    r_c = 1.3 {\rho_c}^{-1/4}\left(1+a_1 \ln(a_2\alpha + 1) + a_3\alpha^{a_4}\right),\label{eq:rcmodel}
\end{equation}

\noindent with $a_1 = -0.25355872$, $a_2 = 0.46241994$, $a_3 = 0.0663722$ and $a_4 = 0.33407792$. Now, for every $\alpha$ there is a $\beta$, which according to the proposed formula (\ref{rho_formula}) fits the density of the solution of  (\ref{eq: GP})-(\ref{eq:poissonSP}), this behavior can be described by the formula

\begin{equation}
    \beta = \frac{b_1\alpha^{b_2}}{\alpha^{b_3} + b_4} + b_5,\label{eq:betamodel}
\end{equation}

\noindent with $b_1 = -1.08334305$, $b_2 = 0.77866182$, $b_3 = 0.81228993$, $b_4 = 6.72089826$ and $b_5 = 1.84588407$. In Figure \ref{fig:rc_beta_fit} we plot the resulting eigenvalue of the system and compare the core radius and beta function against that of the eigen-solution of the eigen-problem (\ref{eq: GP})-(\ref{eq:poissonSP}). The asymptotic behavior of $\beta$ corresponds to the limit in which the solution is the same as the hydrogen atom $\beta \to 1$ when $\alpha \to \infty$.

In order to show these models work and that the  solution of the eigenvalue problem can be fitted with these formulas, we show in Figure \ref{fig:rc_beta_fit} the comparison between the density profiles resulting from the solution of the eigenvalue problem and that obtained with the formulas (\ref{rho_formula}), (\ref{eq:rcmodel}) and (\ref{eq:betamodel}) for $r_c$ and $\beta$ above, for three different values of $\alpha = 0,10,100$. 

  
 {\it Evolution of stationary solution.} 
Now that we have a model for the density profile of a stationary solution we test the stability of configurations constructed in this way. Since the formula does not exactly give the profile obtained from the solution of the eigenvalue problem, the one constructed with the formula can be considered as a solution to the eigenvalue problem plus a perturbation. Then, the effects of the perturbation will trigger oscillation modes of the equilibrium configuration.
 
We now show that these solutions of the SP system with a black hole in the center are stable, and that the code works in this scenario. For this we evolve some configurations with different values of $\alpha$. The evolution is carried out in the numerical  cubic domain of side $L=20$ with the black hole at the center, using resolution $\Delta x=20/128$. The model for the black hole potential uses the prescription in Eq. (\ref{eq:bhmodel}).

\begin{figure}[h]
    \centering
    \includegraphics[width=8cm]{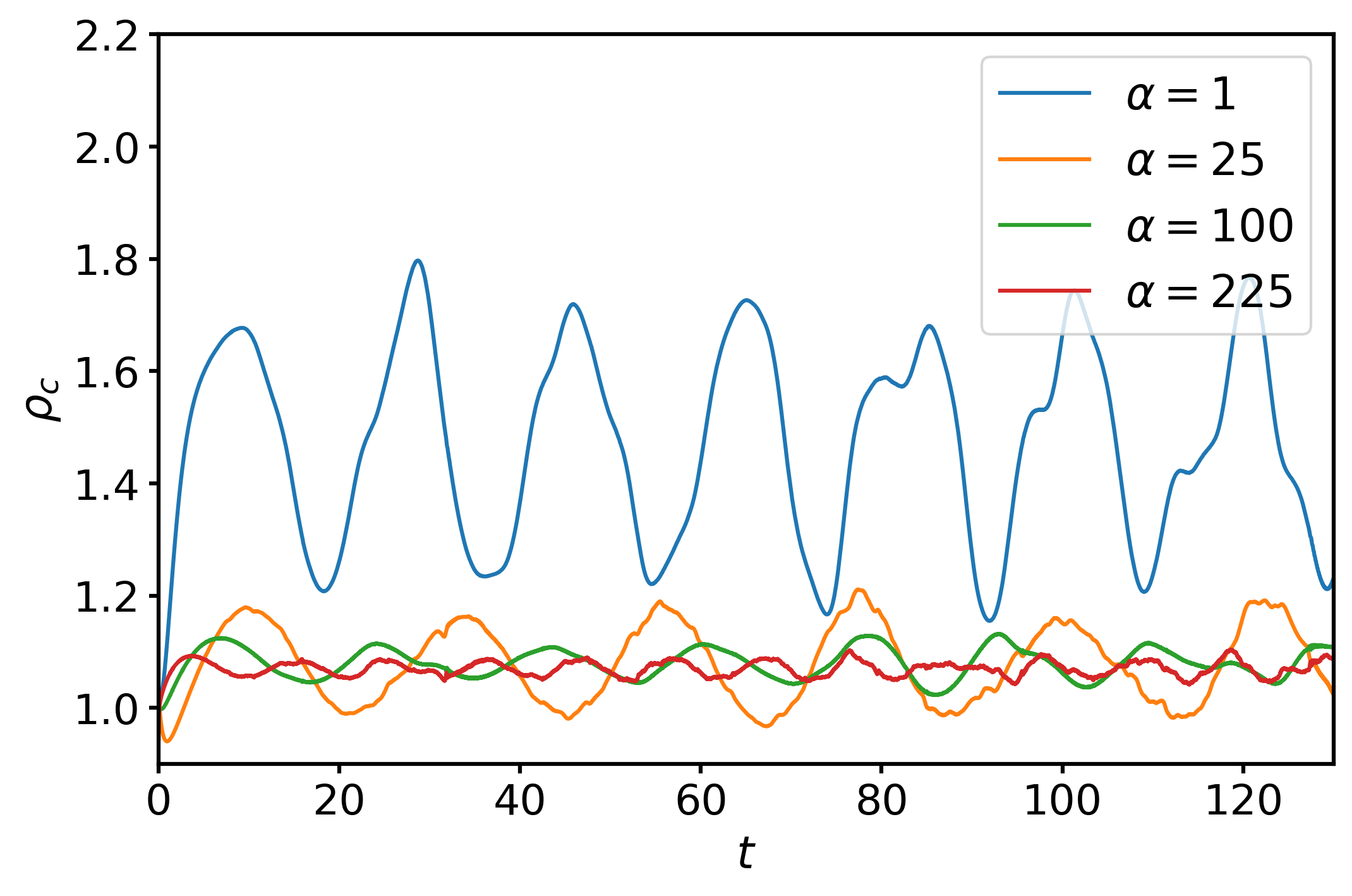}
    \caption{Oscillation of the central density for configurations with different $\alpha$.}
    \label{fig:rhoc_frequencies}
\end{figure}

The results of the evolution of various configurations are shown in Figure \ref{fig:rhoc_frequencies} and can be compared with the numerically perturbed solitons in \cite{GuzmanUrena2006}. Likewise in the pure soliton solution \cite{Guzman2019}, under a perturbation the core oscillates with a dominant frequency mode. 

\section{Test-field condensation of FDM}
\label{app:testfield}

Prior to the main scenario involving an interacting BH with a fluctuating medium of FDM, there is an intermediate scenario in which the BH remains fixed and does not respond to the fluctuating gravitational field sourced by the FDM. This allows us to track the condensation process due to the sole presence of the black hole, without reheating its motion and without the BH scattering the FDM around.

We use the same initial conditions and black hole masses as in the fully coupled case. The evolution is illustrated in Figure \ref{fig:evolution_128_64_32}. 
The condensation process is monitored with the maximum density as function of time in Figure \ref{fig:rhomaxMBHtestfield}. This result is to be contrasted with that in Figure \ref{fig:rhomax_moving}.

\begin{figure}[h]
    \centering
    \includegraphics[width=8.5cm]{Fig11.png}
    \caption{Snapshots of the density at different times for $M_{BH} = M/256,~M/128,~M/64,~M/32$, before and after the black hole captures the minicluster. }
    \label{fig:evolution_128_64_32}
\end{figure}

\begin{figure}[h]
    \centering
    \includegraphics[width=8.5cm]{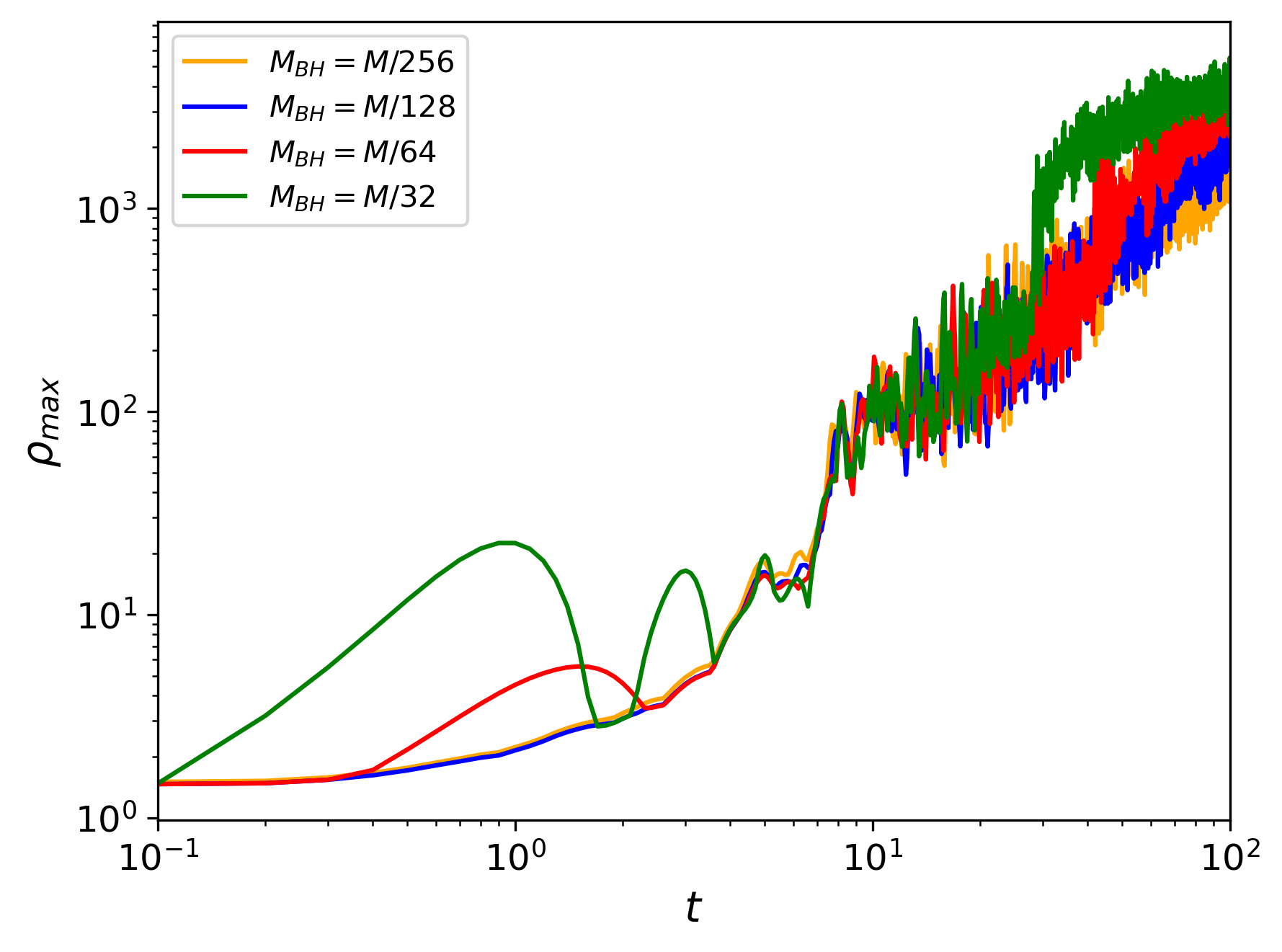}
    \caption{ Evolution of $\rho_{max}$ in simulations with  $M_{BH}=M/256,M/128,M/64,M/32$ in the test field regime. Unlike in the case where the BH reacts to the FDM gravitational potential, the maximum density does not decrease.}
    \label{fig:rhomaxMBHtestfield}
\end{figure}

\end{document}